\RequirePackage{fix-cm}
\documentclass[twocolumn]{svjour3}          
\smartqed  
\usepackage{graphicx}
\usepackage{amsfonts}
\begin{document}

\title{Observation of dipolar transport in one-dimensional photonic lattices
}


\author{Camilo Cantillano\and Luis Morales-Inostroza\and Basti\'an Real\and Santiago Rojas-Rojas\and Aldo Delgado\and Alexander Szameit\and Rodrigo A. Vicencio}


\institute{Camilo Cantillano\and Luis Morales-Inostroza\and Basti\'an Real\and Rodrigo A. Vicencio \at Departamento de F\'{\i}sica, MSI-Nucleus on Advanced Optics, and Center for Optics and Photonics (CEFOP), Facultad de Ciencias, Universidad de Chile, Santiago, Chile\\\email{rvicencio@uchile.cl} \and Santiago Rojas-Rojas\and Aldo Delgado \at Center for Optics and Photonics and MSI-Nucleus on Advanced Optics, Universidad de Concepci\'{o}n, Casilla 160-C, Concepci\'{o}n, Chile \and Alexander Szameit\at Institute for Physics, University of Rostock, Albert-Einstein-Strasse 23, 18059 Rostock, Germany}

\date{Received: date / Accepted: date}

\maketitle

\begin{abstract}

We experimentally study the transport properties of dipolar and fundamental modes on one dimensional (1D) coupled waveguide arrays. By carefully modulating a wide optical beam, we are able to effectively excite dipolar or fundamental modes to study discrete diffraction (single-site excitation) and gaussian beam propagation (multi-site excitation plus a phase gradient). We observe that dipolar modes experience a larger spreading area due to an effective larger coupling constant, which is found to be more than two times larger than the one for fundamental modes. Additionally, we study the effect of non-diagonal disorder and find that while fundamental modes are already trapped on a weakly disorder array, dipoles are still able to propagate across the system.

\end{abstract}

\section{Introduction}
\label{intro}

Waveguide arrays and photonic lattices are an important field of study where many fundamental and applied problems can be investigated in a rather simple configuration \cite{rep1,rep2}. Most of the theoretical and experimental efforts have been focused on studying transport and localization properties in various contexts, such as complex beam steering \cite{ace1,Pertsch1,Silberberg1}, Bloch oscillations \cite{Pertsch2,Peschel1,Trompeter1}, dynamic localization \cite{Longhi1,Szameitdl}, relativistic emulations \cite{Dreisow}, discrete solitons \cite{chris,ds1d,ds2d}, and many more. Recently, even the absence of transport and linear localization in complex lattice geometries was investigated \cite{Liebus,LiebThom,diamond,sawtooth,chenlieb,chenkagome}. 

Importantly, almost all previous works have considered single-mode waveguides only. This somehow reduces the complexity of the studied problem, allowing a more direct verification of theoretical results on simpler experimental setups. But, optical waveguides could also host higher order modes. Their excitation could promote richer dynamics and new interesting phenomena, as it has been suggested for cold-atoms loaded in optical potentials \cite{bec1,bec2,bec3,ref2a} (in that context, dipolar modes are known as p-modes). However, in general, a precise excitation of different modes or complex spatial structures may be simpler using light than atoms \cite{Liebus}, as we will show along this work.

\begin{figure}[h!]
\centering
\includegraphics[width=0.48\textwidth]{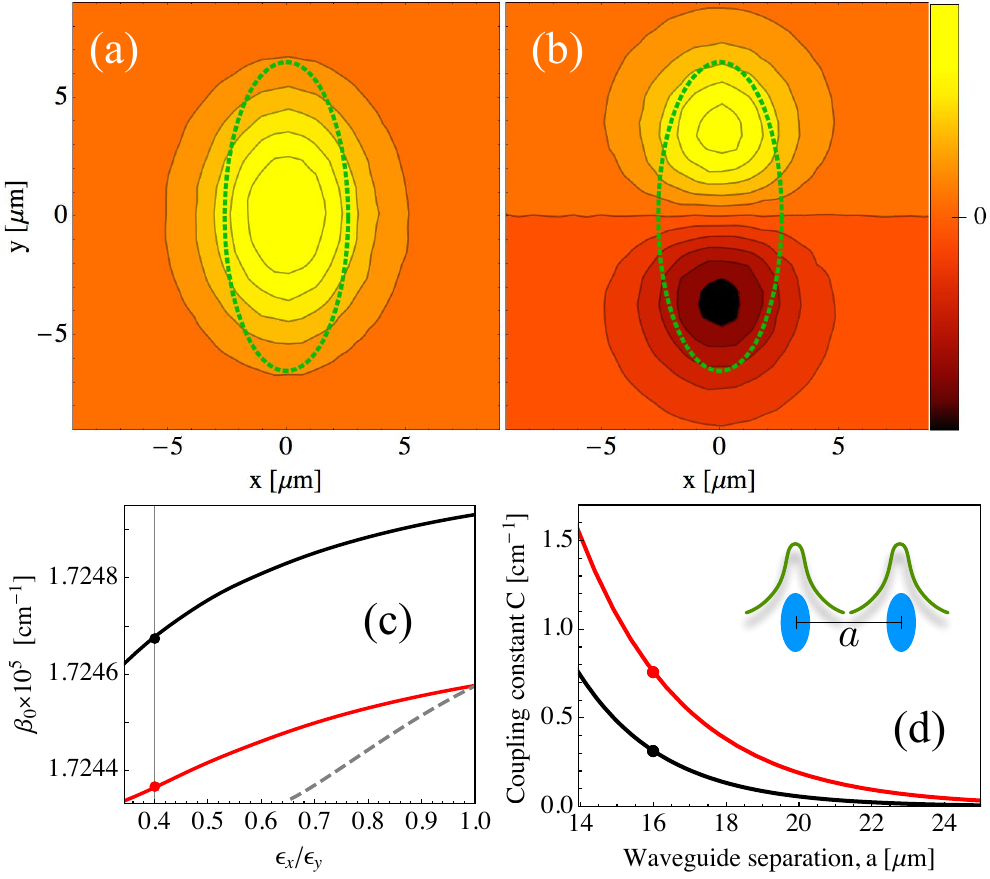}
\caption{Waveguide modes and their constants. {\bf a} Fundamental and {\bf b} dipolar mode profiles of an elliptical waveguide (dotted ellipses indicate the waveguide profile). {\bf c} Solutions diagram in terms of the waveguide geometry $\epsilon_x/\epsilon_y$ and the longitudinal propagation constant $\beta_0$. {\bf d} Coupling constants versus waveguide separation. In {\bf c} and {\bf d} fundamental and dipolar modes are shown in black and red color, respectively.}\label{f1}
\end{figure}

In this work, we present a first systematic study on the diffraction properties of dipolar modes in coupled waveguide arrays. We find that dipoles form another first tight-binding band, that is fundamentally distinct from the higher-order bands of continuous periodical systems \cite{mandelik}. Our waveguide arrays are fabricated using a very precise femtosecond-laser technique \cite{fst}, which produces micrometer waveguides disposed on a given two-dimensional transversal pattern. Light propagating on these waveguides is well trapped in space, allowing a theoretical description based on coupled-mode theory, due to the weak coupling interaction between neighboring waveguides. By using a green laser beam and a modulation setup, we are able to effectively excite fundamental and dipolar modes, and study their dynamical properties in ordered as well as in disordered waveguide lattices \cite{pradiso}.


\section{Waveguide modes}
\label{se2}

A single-mode waveguide could become multimode when reducing the laser wavelength, or when increasing its cross-section or refractive index contrast \cite{snyder,snyder2}. A first excited mode is denominated ``dipolar'' LP11 mode \cite{gloge}, which can have an horizontal or vertical distribution, depending on the particular waveguide geometry. The experimental excitation of higher-order modes has already being reported in Ref. \cite{hexacolor} for highly elliptical two-dimensional waveguides \cite{fst}, although a systematic study in the context of weakly coupled systems is still elusive. 

The analytical treatment for finding the modes of elliptical waveguides is not trivial \cite{ew1}, essentially due to their geometry and complex refractive index profiles. Therefore, we implement a numerical finite-difference method to find the propagating modes of our elliptical waveguides of geometry $\epsilon_x\times \epsilon_y=5.2\ \mu$m $\times 13\ \mu$m, excited using a green laser of $532$ nm. Waveguide parameters were tuned in order to match our experimental observations, using a bulk fused-silica index $n_{0}=1.46$ and a maximum index contrast of $\Delta n=0.73\times 10^{-3}$. [This is obtained by fitting the experimental index profiles from Ref. \cite{fst}, and constructing a continuous index gradient function, where $\epsilon_x$ and $\epsilon_y$ describe the widths of the ellipse shown in Figs.~\ref{f1}(a) and (b)]. We look for modes of a single waveguide and find that, in our highly elliptical regime ($\epsilon_x/\epsilon_y=0.40$), there are only two possible solutions: the fundamental mode sketched in Fig.~\ref{f1}(a) and the vertical dipolar mode shown in Fig.~\ref{f1}(b) [corresponding to dots shown in Fig.~\ref{f1}(c)]. We notice that the fundamental mode is well trapped at the waveguide center, while the dipolar mode presents a more extended tail in the upper and lower region, with zero amplitude at the center. 

By tuning the ratio $\epsilon_x/\epsilon_y$, we find the possible solutions as a function of their longitudinal propagation constant $\beta_0$, as shown in Fig.~\ref{f1}(c). We find that an additional horizontally oriented dipolar mode appears, as shown by a dashed line in Fig.~\ref{f1}(c). This occurs when the waveguide geometry tends to a circular limit ($\epsilon_x/\epsilon_y=1$), where both, horizontal and vertical, dipolar modes converge and degenerate.

Following the method described in Ref. \cite{pradiso}, we computed the horizontal coupling coefficients for fundamental and dipolar modes, for two identical waveguides separated --center to center-- by a given distance $a$ [see Fig.~\ref{f1}(d)]. First of all, we observe a typical exponential decaying tendency for the coupling constant of both modes \cite{fst,hexacolor}. Then, we clearly see that the dipolar horizontal coupling is always larger due to the more extended dipole tail. As an example, for a distance $a=16\ \mu$m, the couplings are $C^f=0.314$ cm$^{-1}$ and $C^d=0.760$ cm$^{-1}$, for the fundamental (f) and dipolar (d) modes, respectively [as indicated by dots in Fig.~\ref{f1}(d)]. [It is important to mention that in the configuration explored along this work (i.e., vertically oriented elliptical waveguides), there is no coupling between fundamental and dipolar modes. This is due to an exact cancellation of the superposition integral for any waveguide separation.]

\begin{figure}[h!]
\centering
\includegraphics[width=0.47\textwidth]{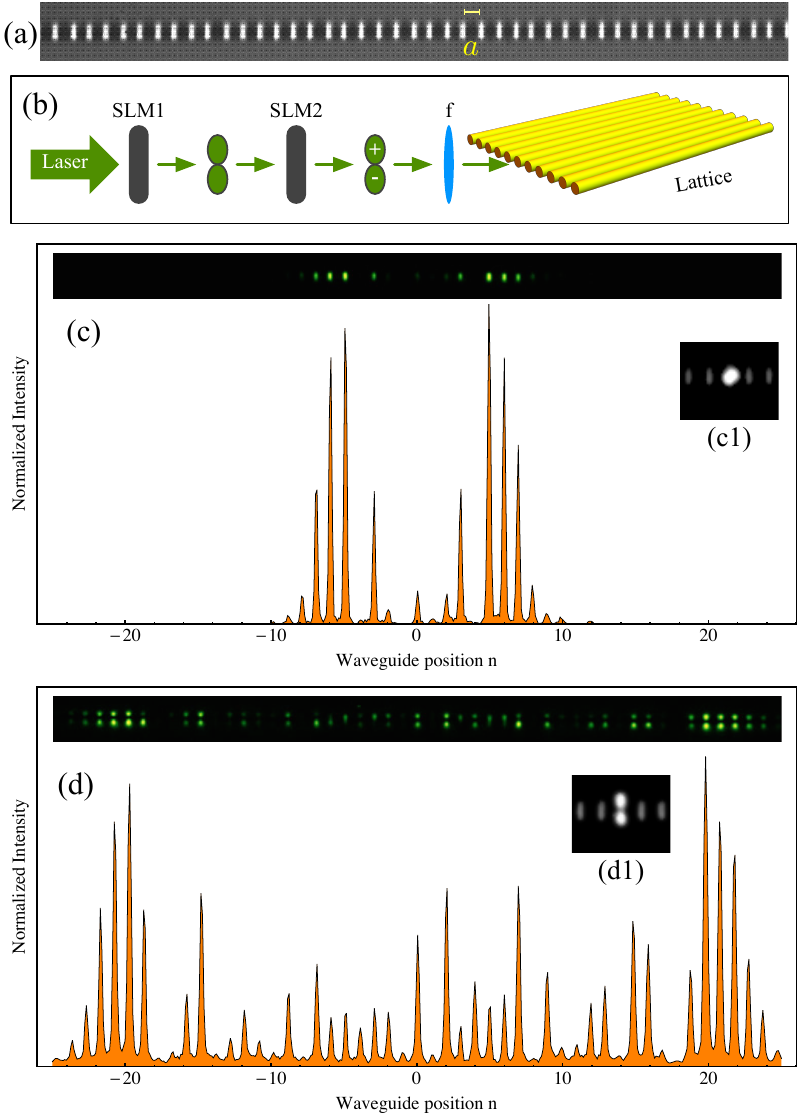}
\caption{Discrete diffraction. {\bf a} Microscope image at the output facet of an ordered 1D photonic lattice. {\bf b} Experimental setup. Discrete diffraction for a single-site {\bf c} fundamental and {\bf d} dipolar mode excitations, using the input profile shown in {\bf c1} and {\bf d1}, respectively. In {\bf c} and {\bf d} an output facet image (top) and an integrated transversal profile (center) are shown.}\label{f2}
\end{figure}

\section{Transport on a 1D lattice}
\label{se3}

We focus on a one-dimensional array of identical elliptical waveguides, as shown in Fig.~\ref{f2}(a) [each white region corresponds to the experimental propagating profile, after white-light illumination]. In this configuration, light trapped at each site of the array interacts only weakly with their surrounding via nearest-neighbor interactions. We describe the dynamics across the lattice using a set of coupled-mode equations \cite{rep1,rep2,Marcuse},
\begin{equation}\label{DLS}
-i \frac {d \psi_n^j}{d z} =\beta_0^j \psi_{n}^j+\left(C^j_{n+1} \psi_{n+1}^j+C^j_{n} \psi_{n-1}^j\right)\ ,
\end{equation}
where $\psi_n^j$ describes the amplitude of the optical field at the $n$th-site, for the fundamental ($j=f$) or dipolar ($j=d$) modes, while paraxially propagating along the longitudinal coordinate $z$. The coefficients $\beta_0^j$ describe the waveguide longitudinal propagation constants, while the coefficients $C^j_{n}$ correspond to the horizontal coupling coefficients between sites $n$ and $n-1$.

We start our study considering an homogenous ordered lattice such that $C^j_n=C^j$. When injecting light on a single lattice site, a well-known pattern is observed after evolution, the so-called discrete diffraction \cite{rep1,rep2}. Its main feature is to concentrate the energy not at the center (as in continuous diffraction), but at the outside external lobes. This linear problem has a formal analytical solution: $\psi_n^j(z)=\psi_0^j i^n J_n(2C^j z)$, where $J_n$ is the Bessel function of order $n$. This pattern is considered as a main signature for a discrete optical system, when experiencing first band dynamics. 

To experimentally study this, we fabricate an ordered lattice of $81$ waveguides [see Fig.~\ref{f2}(a)] with a lattice constant of $a=16\ \mu$m, on a $L=10$ cm long fused silica chip (the geometrical shape of every waveguide corresponds to a super-gaussian of third order, with a cross section of about $4\times 13\ \mu$m$^2$~\cite{cs}). We study linear propagation using an experimental setup based on a sequence of two Spatial Light Modulators (SLMs) as sketched in Fig.~\ref{f2}(b): we first use a transmission Holoeye LC2012 SLM (SLM1) to create an amplitude profile and, then, we modulate its phase using a reflective phase-only Holoeye PLUTO SLM (SLM2). In this way, we are able to excite a lattice injecting a modulated $532$ nm laser beam on a single (or several) waveguide (s) with an input profile, as shown in Figs.~\ref{f2}(c1) and (d1), for fundamental and dipolar excitation, respectively. The generation of dipolar input profile requires amplitude as well as phase modulation in order to mimic the mode profile shown in Fig.~\ref{f1}(b). However, to obtain the right experimental mode profile is not straightforward. We first inject a basic dipolar profile and experimentally observe the output image. We look for a clear dipole profile located at any waveguide and obtain its shape. Then, we use that shape to create an image in the SLM1 as a new input profile (of course, in the SLM2 we add the respective phase). We inject this new profile in the input facet and observe again the output pattern. We repeat this process up to observing only dipoles at the output profile. This is an experimentally iterative method we developed in this work, that allows us to obtain very precise input excitations. 

Figs.~\ref{f2}(c) and (d) show the experimentally obtained discrete diffraction patterns for both input excitations. We observe that dipolar modes experience a larger spreading area compared to a standard fundamental mode excitation. The dipolar diffraction pattern shows a very broad profile, similar to the one observed for a similar lattice but using infrared light at $800$ nm \cite{disoOE}. Our results clearly show the possibility to excite two very different spatial light distributions by simply changing the input profile. This could be used as a switch between two distinguishable orthogonal states, considering that the coupling between fundamental and dipolar modes is always zero in this geometry, without any hybridization \cite{bec3}.

\begin{figure}[h!]
\centering
\includegraphics[width=0.48\textwidth]{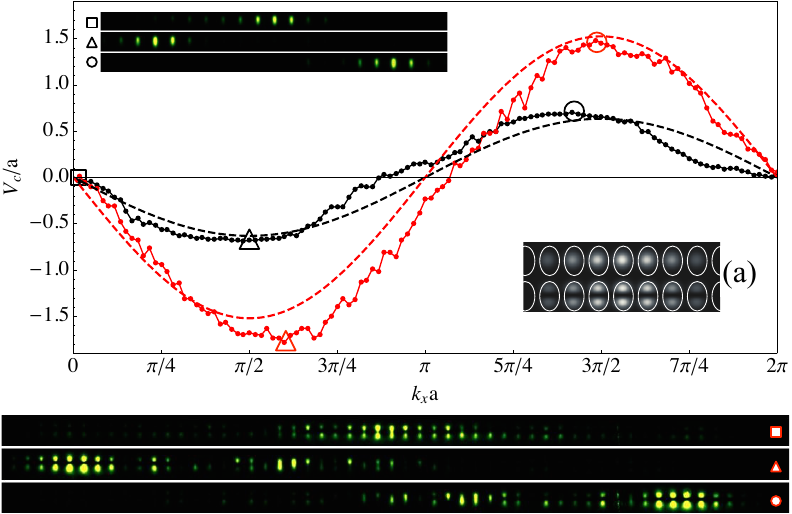}
\caption{Beam propagation. {\bf a} Input profiles. Main figure: Velocity $V_c/a$ of a discretized Gaussian beam versus the normalized transversal wavevector $k_x a$. Dots connected by lines and dashed lines correspond to the experimental and the theoretical data, respectively. Black and red color correspond to fundamental and dipolar modes. Insets show output intensity profiles corresponding to symbol positions.}\label{f3}
\end{figure}

Propagating stationary solutions of model (\ref{DLS}) are obtained using the plane wave (PW) ansatz $\psi_n^j(z)=\psi_0^j\exp{(ik_x n a)}\exp{(i\beta_j z)}$ \cite{Pertsch1,Silberberg1}. We find the system's longitudinal frequencies $\beta_j$ as a function of the transversal wave-vector $k_x$: $\beta_j(k_x)=\beta_0^j+2C^j \cos(k_x a)$. This expression defines two similar linear bands, but shifted in frequency depending on the specific coefficients. Both modes experience first-band dynamics, but in two completely independent bands. $\beta_j(k_x)$ corresponds to the dispersion relation for the lattice modes, and the derivative with respect to $k_x$ gives the transversal discrete PW velocity
\begin{equation}
V_x^j/a\equiv\frac{\partial \beta_j}{\partial k_x}=-2C^j \sin (k_x a)\ ,
\label{vel}
\end{equation}
which becomes zero for $k_x a=m \pi$, with $m\in \mathbb{Z}$. This velocity finds a maximum $|V_x^j/a |=2C^j$ for $k_x a=(2n+1)\pi/2$, with $n\in \mathbb{Z}$. As the linear band is bounded, there is a maximum transversal velocity determined by the coupling coefficients of each excited mode [in fact, the external propagating lobes in Fig.~\ref{f2}(c) and (d) propagate approximately at this maximum velocity, defining the maximum covered area for linear transport on a given lattice]. In order to test this prediction, we take advantage of the capability of our experimental setup and investigate the propagation properties of an ordered lattice by injecting a tilted gaussian beam. This gaussian profile requires to be as wider as possible in order to closely represent a PW of single wave-vector $k_x$. However, real setups are finite in the number of waveguides as well as in the propagation coordinate. Therefore, we implement our experiment using gaussian profiles that cover only $7$ sites of the array, and adjusting the gaussian width to better match the theory (\ref{vel}). Using our SLM setup, we generated discretized gaussian profiles composed of fundamental or dipolar modes, as shown in Fig.~\ref{f3}(a). We made a fine sweep of the input tilt by varying the input phase $\phi=k_x a$ in the interval $\{0,2\pi\}$, with step size of $\pi/60$. For both mode configurations, we took $120$ output profiles at $z=L$ and computed their center of mass transversal velocity, defined as $V_c^j\equiv X_c^j/L$, where $X_c^j=a n_c^j$ and $n_c^j= \sum_n n |\psi_n^j|^2/\sum_n |\psi_n^j|^2$. We collect our experimental results in Fig.~\ref{f3}.
%

We observe a good agreement between the experimental data (dots connected by lines) and the theoretical prediction for the transversal velocity (dashed lines). We made a fit of our experimental data and the theoretical formula (\ref{vel}), obtaining the coupling coefficients: $C^f=0.316$ cm$^{-1}$ and $C^d=0.761$ cm$^{-1}$ (these values are almost equal to the numerical coefficients described before). In the examples shown in Fig.~\ref{f3}, we observe a more visible dispersion for the dipolar gaussian beam compared to the fundamental mode one. This is essentially originated due to the complexity of generating a modulated dipole gaussian profile. But, nevertheless, the center of mass velocity follows a clear sine function tendency, validating our experimental method. Additionally, considering also the discrete diffraction results, we experimentally validate the use of simple (first band) tight-binding models (\ref{DLS}) to theoretically study dipolar excitations on 1D lattices configurations.

\section{Transport on disordered 1D lattices}
\label{se4}

\begin{figure}[h]
\centering
\includegraphics[width=0.48\textwidth]{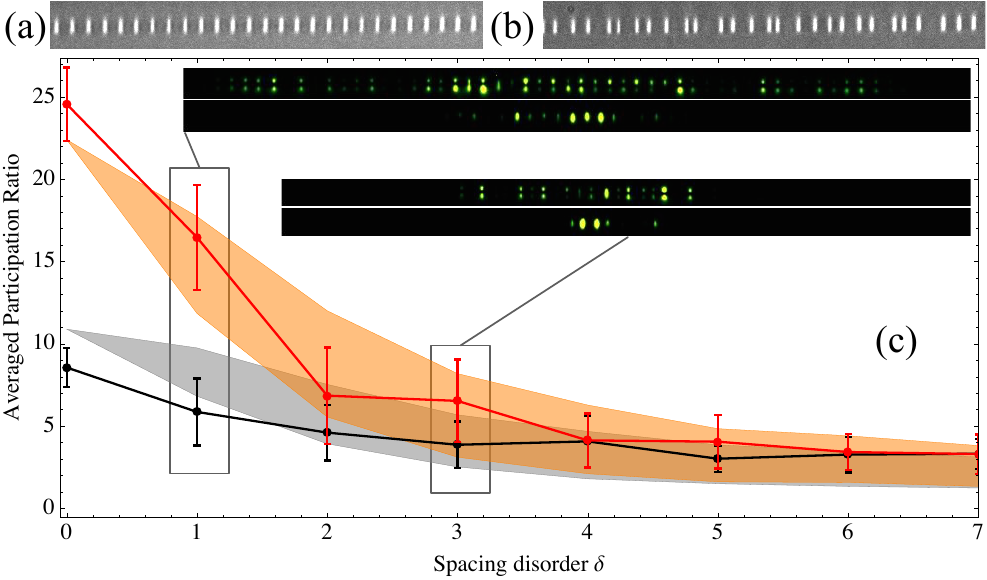}
\caption{Transport in disordered lattices. Microscope image at the output facet of an {\bf a} ordered and {\bf b} disordered one-dimensional photonic lattice. {\bf c} Averaged participation ratio $\bar{R}$ versus spacing disorder $\delta$. Dots show experimental values for fundamental (black) and dipolar (red) modes (bars indicate the experimental standard deviation). The shaded regions show the numerical results for the fundamental (gray) and the dipolar (orange) modes. Insets show two examples.}\label{f4}
\end{figure}

Finally, we study the effect of disorder on 1D waveguide arrays, using fundamental and dipolar excitations. It is well known that disorder induces localization due to multiple destructive interference of randomly distributed scatters \cite{ande1}, what has been already confirmed experimentally in photonic lattices \cite{ande2,ande3}. We fabricated eight lattices with $81$ sites each, where disorder was created by randomly varying the distance between neighboring waveguides in the interval $a\in\{16- \delta,16+ \delta\}\ \mu$m, with $\delta=(0,1,2,3,4,5,6,7)$. $\delta$ is defined as the spacing disorder (a larger $\delta$ implies a larger range of possible distances between neighbor waveguides, therefore an increasing degree of disorder). Our lattices, composed of identical waveguides, present only coupling (off-diagonal) disorder in model (\ref{DLS}); i.e., the coefficients $C^j_n$ are not constant due to the randomness in the horizontal waveguide positions \cite{disoOE,pradiso} [a different waveguide separation produces a different local coupling coefficient between two neighboring waveguides, as expected from Fig.\ref{f1}(d)]. Figs.~\ref{f4}(a) and (b) show examples of an ordered ($\delta=0$) and a disordered ($\delta=5$) lattice. In order to have statistic, we illuminated every array in $40$ different sites using single-site fundamental or dipolar mode excitations [as shown in Figs.~\ref{f2}(b1) and (c1)]. We obtained $40$ output images for every array, and every mode, and computed the respective participation ratio $R^j\equiv(\sum |\psi_n^j|^2)^2/\sum |\psi_n^j|^4$. We obtained an averaged value $\bar{R}^j$, including its standard deviation $\sigma_R$, as shown in Fig.~\ref{f4}(c). We observe an overall tendency to localization for both modes as the disorder strength increases, as expected for disordered finite lattices \cite{disoOE,pradiso}.

%

We numerically integrated model (\ref{DLS}) by considering a random distribution of coupling constants $C_n^j$, in the interval $\{C^j(a+\delta),C^j(a-\delta)\}$. We considered the same range of distances as in the experiments (determined by parameter $\delta$), assuming the dependence of coupling constants presented in Fig.~\ref{f1}(d). We generated $100$ realizations for every value of $\delta$, and obtained the region $\bar{R}^j\pm \sigma_R$, which is shown by shaded areas in Fig.~\ref{f4}(c). First of all, we find a very good qualitative agreement between our experimental and numerical results, validating again the utilization of model (\ref{DLS}) to describe the dynamics of fundamental and dipolar modes on 1D lattices. We observe that the fabricated disordered lattices rapidly conduce to localization for the fundamental mode, while there is still a good transport for dipolar propagation [see examples at $\delta=1$]. For even stronger disorder, dipolar modes still have the opportunity to explore the lattice and disseminate the energy, while the fundamental excitation is already well localized in space. When disorder is very strong ($\delta>4$), both modes tend to spatially localize with an almost equal averaged participation ratio of $\bar{R}\approx 3$. Although there is a strong propagation difference for zero, weak and intermediate disorder, for stronger one any input excitation will remain localized as originally predicted in Ref. \cite{ande1}.

\section{Conclusions}
\label{se5}

In conclusion, we have theoretically and experimentally studied a 1D waveguide array by considering fundamental and dipolar mode excitations. We have shown, using single-site and gaussian beam excitations, that the spreading area is enhanced for dipolar modes in this lattice. Additionally, we have explored the effect of considering disorder on a 1D lattice and have shown that its effect is weaker for dipolar modes, although for stronger disorder both modes localize. After three different experiments, we validate the use of model (\ref{DLS}) as a good theoretical description to study the dynamics of fundamental and dipolar modes in a first-band environment. Extension to hybrid interactions, higher dimensions, and nonlinear effects \cite{bec1,bec2,bec3,magnusdipole,conforti,unisign,denz07,chen2015} are interesting extensions to be explored in detail in the future. Our experimental results may open up a new window to perform p-orbital quantum simulations using photons. Our setup may also provide a more controllable platform for the study of exotic p-orbital phases, which have been previously suggested in the context of cold atoms in optical lattices~\cite{ref2a}.

\begin{acknowledgements}
Authors want to acknowledge M. Johansson for useful discussions and U. Naether for fabricating the lattices. This work was supported in part by Programa ICM grant RC130001, FONDECYT Grant No. 1151444, the Deutsche Forschungsgemeinschaft (grant NO 462/6-1, SZ 276/7-1, SZ 276/9-1, BL 574/13-1), and the German Ministry of Education and Research (Center for Innovation Competence program, grant 03Z1HN31).
\end{acknowledgements}

\end{document}